\documentclass[sigconf]{acmart}

\AtBeginDocument{%
  }

\setcopyright{acmlicensed}
\copyrightyear{2025}
\acmYear{2025}
\acmDOI{XXXXXXX.XXXXXXX}

\acmConference[arXiv'25]{}{}{}
\acmISBN{978-1-4503-XXXX-X/2018/06}

\usepackage{caption}
\usepackage{multirow}
\usepackage{bm}
\usepackage{footmisc}

\begin{document}

\title{Intent-Interest Disentanglement and Item-Aware Intent Contrastive Learning for Sequential Recommendation}

\author{Yijin Choi}
\orcid{0009-0000-2827-7296}
\affiliation{%
  \institution{UNIST}
  \city{Ulsan}
  \country{Republic of Korea}
}
\email{yjchoi0315@unist.ac.kr}

\author{Chiehyeon Lim}
\orcid{0000-0001-6112-9674}
\authornote{Corresponding author}
\affiliation{%
  \institution{UNIST}
  \city{Ulsan}
  \country{Republic of Korea}
}
\email{chlim@unist.ac.kr}

\renewcommand{\shortauthors}{Choi et al.}

\begin{abstract}
Recommender systems aim to provide personalized item recommendations by capturing user behaviors derived from their interaction history. Considering that user interactions naturally occur sequentially based on users' intents in mind, user behaviors can be interpreted as user intents. Therefore, intent-based sequential recommendations are actively studied recently to model user intents from historical interactions for a more precise user understanding beyond traditional studies that often overlook the underlying semantics behind user interactions. However, existing studies face three challenges: 1) the limited understanding of user behaviors by focusing solely on intents, 2) the lack of robustness in categorizing intents due to arbitrary fixed numbers of intent categories, and 3) the neglect of interacted items in modeling of user intents. To address these challenges, we propose Intent-Interest Disentanglement and Item-Aware Intent Contrastive Learning for Sequential Recommendation (IDCLRec). IDCLRec disentangles user behaviors into intents which are dynamic motivations and interests which are stable tastes of users for a comprehensive understanding of user behaviors. A causal cross-attention mechanism is used to identify consistent interests across interactions, while residual behaviors are modeled as intents by modeling their temporal dynamics through a similarity adjustment loss. In addition, without predefining the number of intent categories, an importance-weighted attention mechanism captures user-specific categorical intent considering the importance of intent for each interaction. Furthermore, we introduce item-aware contrastive learning which aligns intents that occurred the same interaction and aligns intent with item combinations occurred by the corresponding intent. Extensive experiments conducted on real-world datasets demonstrate the effectiveness of IDCLRec.
\end{abstract}




\maketitle

\section{Introduction} 

Recommender systems are widely utilized in various fields such as e-commerce services to enhance user experiences by providing personalized item recommendations~\cite{lu2015recommender, bobadilla2013recommender}. The primary goal of recommender systems is to accurately predict users' preferences by leveraging their past interactions based on the assumption that their preferences can be inferred from historical data. Given that user interactions occur inherently sequential, many sequential recommendation models are developed to capture the temporal patterns of user behaviors. These models are designed to predict the next item that the user will interact with and have demonstrated good performance in recommendation tasks~\cite{kang2018self, xie2022contrastive, tang2018personalized, chang2021sequential, fan2022sequential}.

Traditional sequential recommendation models focus primarily on capturing behavioral patterns of users without delving into the underlying semantics behind the behaviors. Recently, considering that users make interactions under their intents in mind, intent-based sequential recommendations are actively studied with emphasizing the understanding of user behaviors from the perspective of intrinsic intentions~\cite{jannach2024survey, chen2022intent, qin2024intent, tanjim2020attentive, li2021intention}. In these approaches, user intents generally refer to the diverse and dynamic motivations that drive users to interact with items. These intents can be changed across various intent categories over time depending on the user's evolving purposes at each time step and the contextual attributes associated with items. Accordingly, these models aim to model intent representations that effectively capture the dynamic nature of user intents as behavior representations.

%\captionsetup[figure]{skip=5pt}
\begin{figure}[t]
\centering
\includegraphics[width=0.99\linewidth]{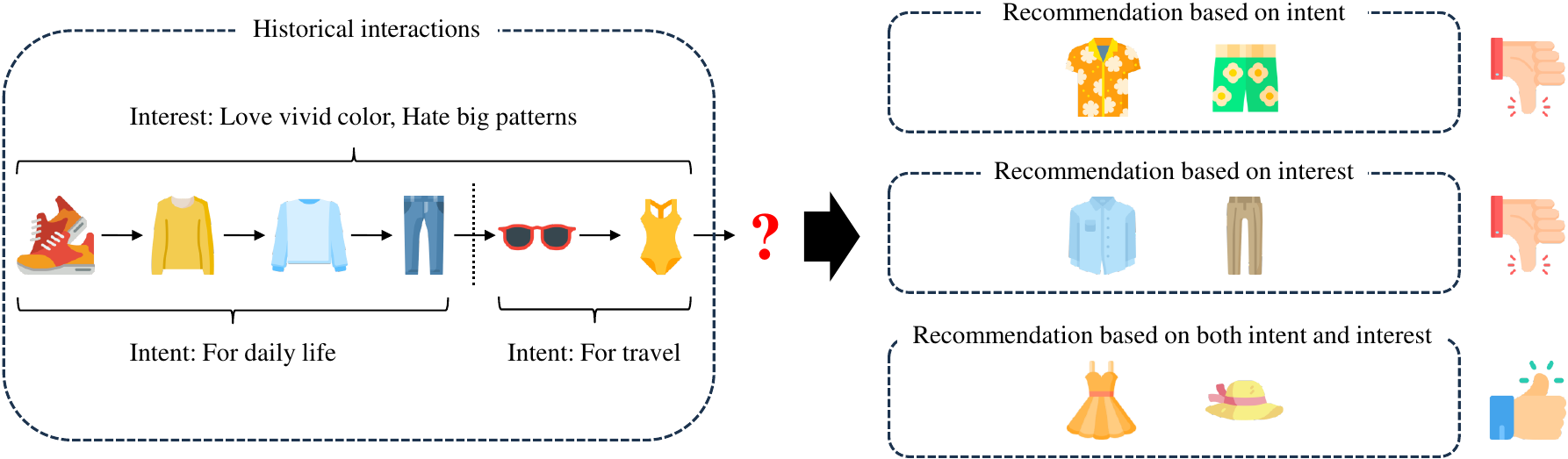} 
\caption{An example of user interactions and corresponding recommendation reflecting different semantic aspects in user behaviors.} \label{example}
%\vspace{-10pt}
\end{figure}

Such studies that interpret user behaviors as intents have shown remarkable performance. However, there are three challenges to be addressed to achieve a more comprehensive understanding of user behaviors and intents. First, focusing only on intentions may result in insufficient understanding of user behaviors and preferences. When users interact with items, they do not simply choose items that align solely with their immediate purposes but also take other factors into account. For instance, consider a user who recently purchased items under the intention of preparing clothes for travel as illustrated in Figure~\ref{example}. Many travel-related items such as floral T-shirts or bright pants can be recommended based on the general purpose of traveling. However, a user can naturally have her own personal tastes for items such that she loves vivid colors while hates big patterns. In this case, recommending floral T-shirts solely based on the travel intent may not be preferable by the user. Consequently, relying exclusively on the intent perspective may fail to fully capture the complexity of user behaviors. Second, explicitly fixing the number of intent categories is not a robust approach for capturing the categorical characteristics of intents. Intents are generally understood as categorical characteristics of user behaviors that apply across all users. Many studies cluster intent representations across all users or calculate the probability of each user’s intents belonging to specific intent categories to divide the overall intents into a predefined number of intent categories. Each intent representation is then aligned with the corresponding cluster or assigned to the category with the highest probability. However, the number of intent categories varies widely across studies, ranging from 3 to 1024~\cite{wang2024intent, li2021intention, chen2022intent, li2023multi, qin2024intent}. This significant variance suggests that arbitrarily dividing intents into a fixed number of categories lacks the robustness to accurately represent categorical intents. Lastly, overlooking interacted items may limit the understanding of user intents. Recent studies leverage contrastive learning to enhance intent representations by aligning augmented intent representations or bringing them closer to their assigned intent categories~\cite{chen2022intent, qin2024intent, liu2024end}. However, since these methods inevitably face the second challenge, these contrastive learning approaches may be ineffective by guiding intents to inaccurate categories. In this situation, even though interacted items that are the only explicit signals of user behaviors could play a crucial role in learning user intents, they are neglected in current studies.

To address these challenges, we propose Intent-Interest Disentanglement and Item-Aware Intent Contrastive Learning for Sequential Recommendation (IDCLRec) which introduces intent-interest disentanglement to tackle the first challenge and item-aware intent contrastive learning to tackle the second and third challenges. When we consider factors other than intents in user behaviors, it becomes evident that user interactions occur through both consistent behaviors related to users' personal tastes across overall interactions in addition to time-varying behaviors associated with specific purposes which is intent~\cite{engel1969personality, belk1975situational}. Although existing studies overlook these consistent properties of user behaviors, we capture them and call as interests. Then, by viewing user behaviors as the combination of intents and interests, we can explicitly disentangle user behaviors into intents and interests that are generally can be seen as long-term and short-term behaviors. Such disentanglement can lead to better understand of more precise behaviors and better learning of intent representations in further contrastive learning. Furthermore, considering the personalized nature of recommendation tasks, we focus on capturing the categorical characteristics of individual user-level intents without predefining the number of intent categories. Given the categorical intent indicating a specific intent category, we incorporate combinations of items interacted with under the same intent category for each user in contrastive learning to provide valuable guidance for capturing nuanced intent representations. The detailed model structure of IDCLRec including these components is compared with existing studies in Table~\ref{modelcomparison}. Specifically, we design a causal cross-attention mechanism that identifies interests from user behaviors by reflecting the consistent characteristics of interests over time. The remaining behaviors after excluding interests are modeled as intents by incorporating the dynamic nature of intents through a similarity adjustment loss function that enforces similarity across adjacent time steps. Also, an importance-weighted attention mechanism is devised to learn categorical intents by integrating intent representations at different time steps that share high similarity with assigning higher importance to more similar intents. To further enhance the derived intent representation, we employ item-aware contrastive learning by considering user intents from both single-item and multi-item combination perspectives. The contrastive objectives align intent representations corresponding to the same target item and ensure that intent representation and item combinations interacted under the same intent category to be closer. Extensive experiments conducted on three real-world datasets demonstrate that IDCLRec significantly outperforms existing models, leading to more personalized and accurate recommendations through intent-interest disentanglement and user specific intent learning.

%\captionsetup[table]{skip=5pt}
\begin{table*}[t]
\renewcommand{\arraystretch}{1.1}
\centering
\caption{Component comparisons of the proposed model with existing models.}
\label{modelcomparison}
\resizebox{0.85\textwidth}{!}{%
\begin{tabular}{c|ccc|c|cc}
\hline
{Model} & \multicolumn{3}{c|}{User Behavior Semantic Perspective} & Intent Separation & \multicolumn{2}{c}{Contrastive Learning Perspective} \\ \cline{2-4} \cline{6-7}
 & \:\: General \quad & \quad Intent \quad & Interest &  & Self-Representation & Cross-Representation \\ \hline\hline
GRU4Rec (2015) & \: \checkmark & \,\, $\boldsymbol{\times}$ & $\boldsymbol{\times}$ & \textbf{--} & \textbf{--} & \textbf{--} \\
SASRec (2018) & \: \checkmark & \,\, $\boldsymbol{\times}$ & $\boldsymbol{\times}$ & \textbf{--} & \textbf{--} & \textbf{--}  \\
BERT4Rec (2019) & \: \checkmark & \,\, $\boldsymbol{\times}$ & $\boldsymbol{\times}$ & \textbf{--} & \textbf{--} & \textbf{--}  \\
S\textsuperscript{3}-Rec\textsubscript{MIP} (2020) & \: \checkmark & \,\, $\boldsymbol{\times}$ & $\boldsymbol{\times}$ & \textbf{--} & \textbf{--} & \textbf{--}  \\
CL4SRec (2022) & \: \checkmark & \,\, $\boldsymbol{\times}$ & $\boldsymbol{\times}$ & \textbf{--} & \checkmark & $\boldsymbol{\times}$  \\
CoSeRec (2021) & \: \checkmark & \,\, $\boldsymbol{\times}$ & $\boldsymbol{\times}$ & \textbf{--} & \checkmark & $\boldsymbol{\times}$  \\
DuoRec (2022) & \: \checkmark & \,\, $\boldsymbol{\times}$ & $\boldsymbol{\times}$ & \textbf{--} & \checkmark & $\boldsymbol{\times}$  \\
DSSRec (2020) & \: $\boldsymbol{\times}$ & \,\, \checkmark & $\boldsymbol{\times}$ & \checkmark & \textbf{--} & \textbf{--}  \\
ICLRec (2022) & \: $\boldsymbol{\times}$ & \,\, \checkmark & $\boldsymbol{\times}$ & \checkmark & \checkmark & $\boldsymbol{\times}$ \\
IOCRec (2023) & \: $\boldsymbol{\times}$ & \,\, \checkmark & $\boldsymbol{\times}$ & \checkmark & \checkmark & $\boldsymbol{\times}$  \\
ICSRec (2024) & \: $\boldsymbol{\times}$ & \,\, \checkmark & $\boldsymbol{\times}$ & \checkmark & \checkmark & $\boldsymbol{\times}$  \\ \hline
IDCLRec & \: $\boldsymbol{\times}$ & \,\, \checkmark & \checkmark & $\boldsymbol{\times}$ & \checkmark & \checkmark  \\ \hline
\end{tabular}%
}
\end{table*}

\section{Related Works} 

\subsection{Sequential Recommendation}

Sequential recommendation aims to predict the next item that user will interact with by capturing the temporal patterns from users' historical interactions~\cite{kang2018self, wang2019sequential, chang2021sequential}. Initial works on sequential recommendation often utilized markov chains that model user transitions between items as probabilistic processes~\cite{he2016fusing}. With the development of deep learning, sequential recommendation models introduced neural networks to extract complex sequential patterns such as recurrent neural networks and convolutional neural networks-based models~\cite{hidasi2015session, tang2018personalized}. In recent years, transformer-based models gained prominence with their powerful performance through attention mechanisms~\cite{kang2018self, sun2019bert4rec, fan2022sequential}. However, these models commonly focus on capturing behavioral patterns of users without considering the underlying semantics behind the behaviors.

\subsection{Intent Recommendation}

Recently, many studies focus on modeling user intents to understand user behaviors from an intent perspective for more effective recommendations~\cite{jannach2024survey, chen2022intent, kang2024hicl, jin2023dual}. These studies can broadly be categorized into two groups which model implicit intent through utilizing side information and which model latent intent only from user interactions. Some studies introduce side information such as action types or item categories by considering that such side information provide valuable signals about user intents of interactions~\cite{tanjim2020attentive, li2022coarse, cai2021category}. However, since side information is not always available, other studies try to model intents by reflecting only interactions~\cite{chen2022intent, tan2021sparse, qin2024intent}. These studies typically assign each intent to a predefined number of intent categories and employ contrastive learning in terms of intents to learn latent intents. Although they can capture intents well, their general approaches are insufficient in fully exploring the diverse characteristics of intents. Moreover, only focusing on intents also can lead to an underexploration of nuanced user behaviors.

\subsection{Contrastive Learning}

Contrastive learning which is a prominent self-supervised learning approach has achieved success in various domains such as computer vision~\cite{chen2020simple, he2020momentum}, natural language processing~\cite{fang2020cert, gao2021simcse}, and recommender systems~\cite{xie2022contrastive, zhou2023equivariant}. It enhances representations learned from unlabeled data by giving signals by maximizing agreement between positive pairs which are typically derived as augmentations of the same data. The sequential recommendation studies with contrastive learning usually seek for effective data augmentation strategies or suitable positive representations to align with~\cite{liu2021contrastive, qin2023meta}. While these approaches show good performance with enhancing representations, they often overlook the signals from items when contrasting representations.

\section{Intent Analyses}

In this paper where user behaviors are represented through both user interests and intents, we conduct intent analyses to capture the characteristics of intents that underlie interactions with items for comprehensive user intent modeling. To achieve this, we analyze intent representations obtained by ICSRec which is a current state-of-the-art (SOTA) model in intent recommendation to check its characteristics. Specifically, ICSRec learns a single unified intent representation for each user from their entire interaction sequence, implying that the entire sequence of interactions reflects a particular user intent. To categorize these intent representations, ICSRec assumes that there are K distinct intent categories applicable to users. Accordingly, it applies K-means clustering to group all users’ intent representations into K clusters. Each cluster ideally corresponds to a particular type of intent containing similar intent representations and the cluster centroid serves as the representative vector for that intent category. Although this clustering does not explicitly provide a semantic label for each category, it can assign each user’s intent representation to a specific category characterized by a vector. Through learned intent representations, we examine the categorical and item-related characteristics of intents to get insights that inform the design of a model capable of more effectively capturing intrinsic intent features.

%\captionsetup[figure]{skip=5pt}
\begin{figure}[b]
\centering
\includegraphics[width=0.99\linewidth]{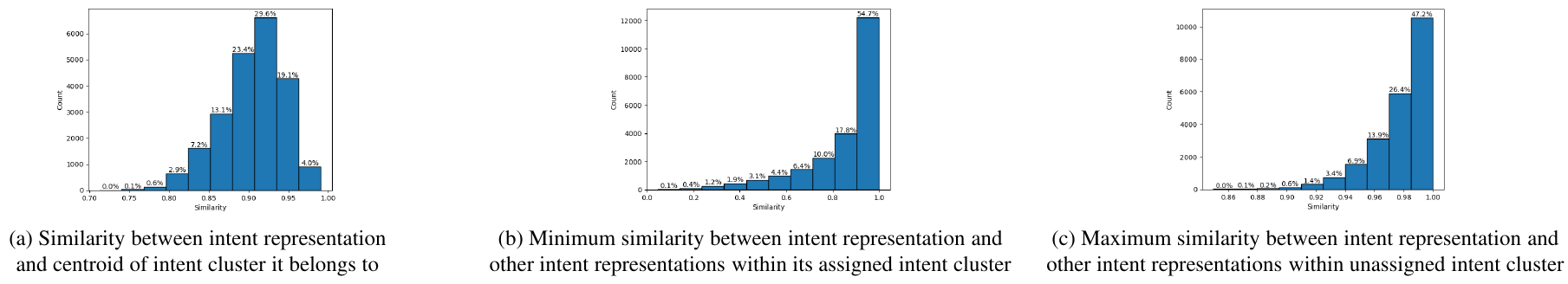} 
\caption{Similarity distributions between intent representations.} \label{analyses}
%\vspace{-10pt}
\end{figure}

As noted in previous works, user intents can be understood as a categorical trait of behaviors. To dive into categorical nature of intents, we analyze the intents derived by ICSRec for Beauty dataset in terms of similarity distributions between intents as shown in Figure~\ref{analyses}. Here, similarity is calculated as a sigmoid of the dot product considering the goal of recommendation task that suggests items with high similarity to users determined by the dot product. Specifically, Figure~\ref{analyses}-(a) shows the distribution of similarity between each intent representation and centroid of cluster it belongs to. Where there are intents with high or relatively low similarity to the centroid of assigned clusters, such variation in similarity indicates that the degree to which each intent contributes to defining the intent category differs. In this context, the intents with higher similarity can be interpreted as more significant representation that define the intent category represented by the assigned cluster. Likewise, the distribution of the minimum similarity between each intent and other intent representations within the same cluster is presented in Figure~\ref{analyses}-(b). The presence of low similarity with other intents for some intents means the existence of dissimilar intents within the same cluster. Such intents might be less important in representing the intent category compared to those with consistently high similarity to other intents. Figure~\ref{analyses}-(c) illustrates the distribution of the maximum average similarity for each intent by computing the average similarity between the intent and all other intents from each different cluster and identifying the highest average similarity among all clusters. This reveals that most intents exhibit high similarity with a specific cluster they do not belong to. Such high similarity suggests that intents share characteristics similar not only to their own intent category but also to other intent categories. Consequently, it can inferred that the commonly used K-means clustering might not be effective in learning intent representations that distinctly capture specific intent categories. In summary, general clustering might not clearly group intents indicating a specific intent category and the reflection of relative importance of intents might be effective to capture intent categories.

%\captionsetup[figure]{skip=5pt}
\begin{figure}[t]
\centering
\includegraphics[width=0.25\linewidth]{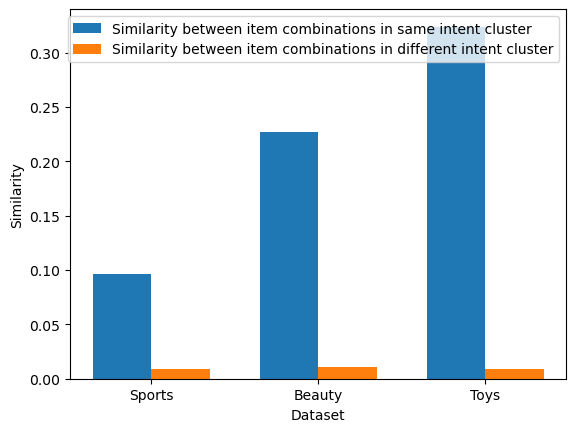} 
\caption{Average similarity between item combinations in same and different intent clusters.} \label{itemsim}
\vspace{-10pt}
\end{figure}

We further investigate the relationship between intents and item combinations by analyzing the item sequences associated with each intent category. Figure~\ref{itemsim} shows the average similarity of item sequences that output intent representations within the same cluster and across different clusters for Sports, Beauty, and Toys datasets. Each item sequence containing the items interacted with by each user is considered to indicate item combination that reflects a specific intent category by deriving a single intent representation. These sequences are represented as simple one-hot encoded vectors which assign 1 to interacted items and 0 to non-interacted items. The similarity between sequences is computed as the number of shared items using the dot product of these vectors. Although the similarity values are generally low due to the highly sparse nature of item sequences from the large number of items, the similarity of item sequences within the same cluster is much higher than the similarity of sequences across different clusters in all datasets. This higher similarity indicates that item sequences associated with similar intents tend to contain more overlapping items. In other words, a relationship exists between item combinations and intents such that item combinations interacted under the same intent category exhibit more similar patterns.

Based on these findings, the following conclusions can be drawn: 1) For the purpose of predicting individual preferences in recommendation tasks, it could be effective to extract similar intents for each user and learn categorical intents by reflecting their relative importance rather than separating all users’ intents. 2) It could be effective to enhance intents in terms of items by considering the item combinations interacted with under users’ current intent category for better alignment of item-level interactions and underlying user intents. Building upon these conclusions, we devise a model that explicitly incorporates modules for learning such beneficial intent characteristics, ensuring that the identified characteristics are effectively reflected in the learned representations.

\section{Methodology} 

\subsection{Problem Definition}

Let $\mathcal{U}$ and $\mathcal{V}$ denote set of users and items, respectively. For each user $u \in \mathcal{U}$, the user interaction history is represented as chronologically ordered sequence of items $S^{u}=[v_1^{u}, v_2^{u}, ..., v_{|S^{u}|}^{u}]$ where $v_t^{u} \in \mathcal{V}$ is the item that user $u$ interacted with at the $t$-th time step and $|S^{u}|$ is the number of interacted items for user $u$. To standardize length of sequences, all user sequences are truncated or padded to satisfy a fixed maximum length $N$. Specifically, if $|S^{u}| \ge N$, the most recent $N$ interacted items are considered as $S^{u}=[v_{|S^{u}|-N+1}^{u}, ..., v_{|S^{u}|}^{u}]$. Otherwise, if $|S^{u}|<N$, we pad the sequence with a special padding item $v_0 \notin \mathcal{V}$ at the beginning until the length of sequence reaches $N$ as $S^{u}=[v_0, ..., v_0, v_1^{u}, ..., v_{|S^{u}|}^{u}]$. The goal of sequential recommendation is to predict the next item $v_{|S^{u}|+1}^{u}$ that user $u$ is most likely to interact with at the time step $|S^{u}|+1$ given historical interaction sequence $S^{u}$, which can be formulated as follows:
\begin{equation}
\underset{v_i \in \mathcal{V}}{\mathrm{argmax}} \; P\left(v_{|S^{u}|+1}^{u}=v_i \mid S^{u}\right).
\end{equation}

\subsection{Overview}

%\captionsetup[figure]{skip=5pt}
\begin{figure*}[t]
\centering
\includegraphics[width=0.9\linewidth]{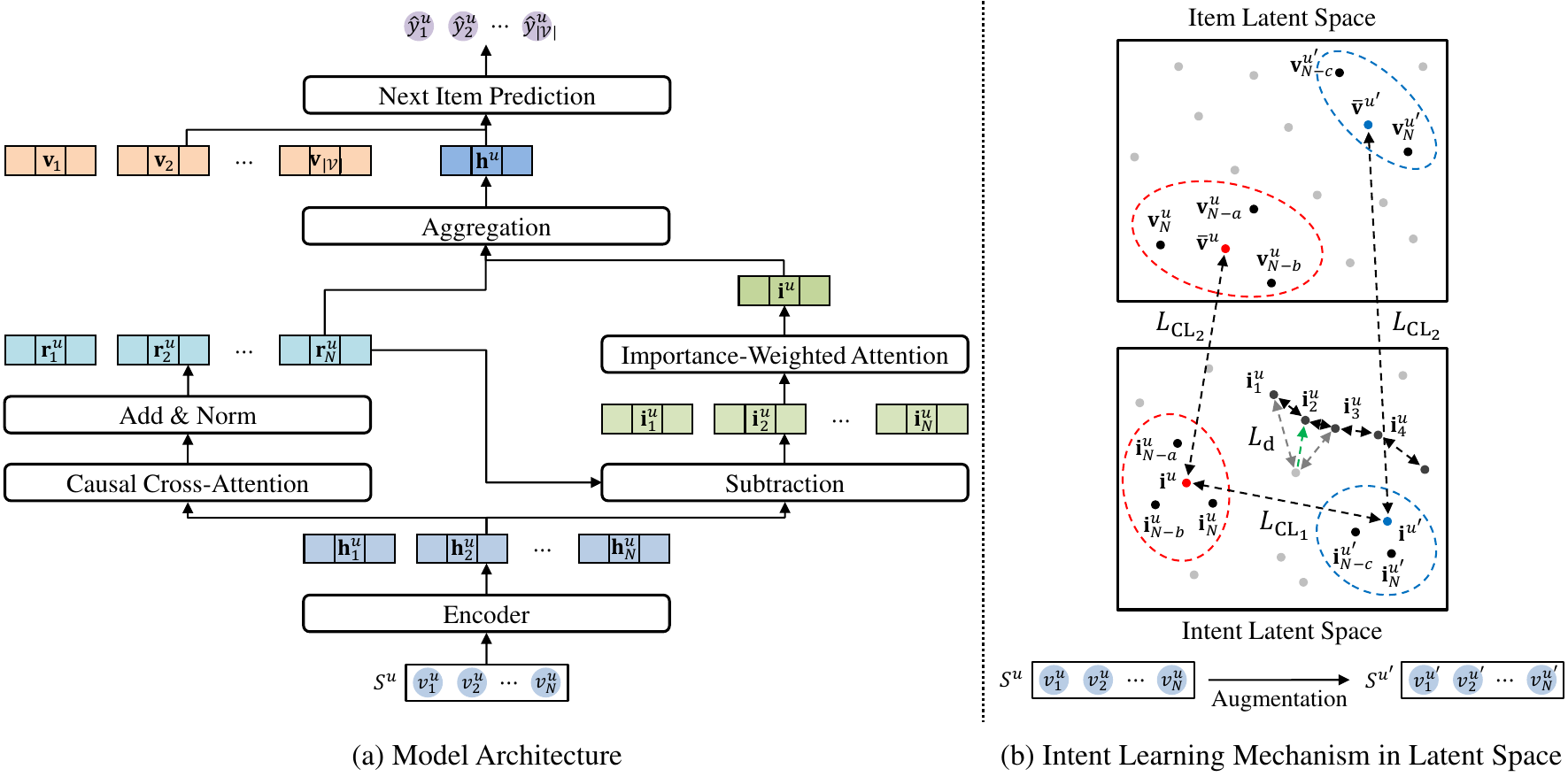} 
\caption{Overview of the proposed IDCLRec. (a) illustrates the architecture of IDCLRec including the modules for each process. (b) presents the mechanism of the training loss function specifically designed for intent representation learning.} \label{overview}
%\vspace{-10pt}
\end{figure*}

The overall framework of IDCLRec is shown in Figure~\ref{overview}. The model aims to fully understand user behaviors by exploring both user interest and intent perspectives. The architecture consists of three principal components: interest representation learning, intent representation learning, and contrastive learning for intent representation. First, the sequence encoder learns user behavior representations from interaction history. To capture interests, we devise a causal cross-attention mechanism that reflects the consistent characteristics of interests, allowing the model to extract interests from sequence representations. We then introduce a loss function to learn intents by focusing on dynamic characteristics. We also perform an importance-weighted attention which leverages the similarity between intents for intent extraction and aggregation to express categorical characteristics of user intents. Finally, two contrastive learning tasks are introduced to further enhance the intent representation in terms of items. The first contrastive task aligns intent representations of sequences sharing the same target item and the second brings the intent representation closer to the items interacted with under the corresponding intent. By combining the learned interest and intent representations, the model predicts the next item a user is likely to interact with.

\subsection{Sequence Representation Learning}

Transformer has been extensively used in sequential recommendation tasks to learn user sequence representations. Especially, SASRec~\cite{kang2018self} stands out as a representative model due to its effective utilization of the transformer encoder architecture. Recent studies have frequently employed SASRec for representation extraction by leveraging its powerful capabilities~\cite{wang2022explanation, zhang2023adaptive, zhou2023equivariant, qin2024intent}. Motivated by these strengths, we adopt SASRec as an encoder for sequence representation learning.

\subsubsection{Embedding Layer}

For the item set $\mathcal{V}$, we construct an item embedding matrix $\mathbf{V} \in \mathbb{R}^{(|\mathcal{V}| + 1) \times d}$ where $d$ represents the dimensionality of the embedding vectors. The additional dimension in $\mathbf{V}$ is allocated for a padding item $v_0$. To incorporate the positional information of items, we build a position embedding matrix $\mathbf{P} \in \mathbb{R}^{N \times d}$. Then, given a user sequence $S^{u}$, sequence embedding $\mathbf{E}^u \in \mathbb{R}^{N \times d}$ is computed by adding the item embedding $\mathbf{V}^u$ with the corresponding position embedding $\mathbf{P}$ as follows:
\begin{equation}
\mathbf{E}^u = \mathrm{Embedding}(S^{u}) = \mathbf{V}^u + \mathbf{P} = \left[ \mathbf{v}_1^u+\mathbf{p}_1, \mathbf{v}_2^u+\mathbf{p}_2, ..., \mathbf{v}_N^u+\mathbf{p}_N \right],
\end{equation}
where $\mathbf{v}_t^u \in \mathbb{R}^d$ is the embedding of item $v_t^u$ which is interacted with at the $t$-th time step in the sequence and $\mathbf{p}_t \in \mathbb{R}^d$ is the positional embedding corresponding to the $t$-th time step.

\subsubsection{Sequence Representation Learning Layer}

After obtaining the sequence embedding, we employ a sequence encoder to learn sequence representation that captures the evolving behavior patterns of user over time. Formally, we define the sequence encoder $f(\cdot)$ which transforms the sequence embedding into a sequence representation $\mathbf{H}^u \in \mathbb{R}^{N \times d}$ as follows:
\begin{equation}
\mathbf{H}^u = f\left(\mathbf{E}^u\right),
\end{equation}
where $f(\cdot)$ denotes the SASRec model in this study. The $\mathbf{H}^u = [\mathbf{h}_1^u, \mathbf{h}_2^u, ..., \mathbf{h}_N^u]$ indicates user behaviors across all time steps where each vector $\mathbf{h}_t^u \in \mathbb{R}^d$ represents the user's behavioral representation at time step $t$, reflecting interaction history until time step $t$.

\subsection{Interest Representation Learning}

User behaviors can be composed of two main components: inherent interests and dynamic intentions. Interests represent unique characteristics of users which tends to remain consistent across different time steps. On the other hand, intentions capture the contextual motivations that can be changed at each time step. To disentangle these components from the user behaviors, we focus on their distinct properties. Specifically, recognizing that interests emerge as stable user behavior patterns across the user interaction history, we introduce a causal cross-attention mechanism to extract interests from the user behaviors.

\subsubsection{Causal Cross-Attention Mechanism}

The causal cross-attention mechanism aims to learn interest representation by focusing on temporal dependencies in user behaviors. By extracting similar information between the user's current and previous interaction behaviors, the model captures consistent patterns across user behaviors.

First, we construct a padded version of the original sequence representation $\mathbf{H}^u$ denoted as $\mathbf{H}^u_{\text{pad}} = [\mathbf{\bar{0}}, \mathbf{h}_1^u, \mathbf{h}_2^u, ..., \mathbf{h}_N^u]$ where $\mathbf{\bar{0}} \in \mathbb{R}^d$ is a zero vector for padding. The user behaviors at the current and previous time steps are represented as $\mathbf{H}^u_{\text{cur}} = [\mathbf{h}_1^u, \mathbf{h}_2^u, ..., \mathbf{h}_N^u]$ and $\mathbf{H}^u_{\text{pre}} = [\mathbf{\bar{0}}, \mathbf{h}_1^u, ..., \mathbf{h}_{N-1}^u]$, respectively. To capture similarities between these behaviors, we employ a causal cross-attention mechanism based on scaled dot-product attention~\cite{vaswani2017attention} which is formulated as:
\begin{equation}
\mathbf{\hat{R}}^u = \mathrm{Attention}(\mathbf{Q}_1, \mathbf{K}_1, \mathbf{V}_1) = \mathrm{softmax} \left(\frac{\mathbf{Q}_1 \mathbf{K}_1^{\mathrm{T}}}{\sqrt{d}} + \mathbf{M} \right)\mathbf{V}_1.
\end{equation}
$\mathbf{Q}_1 = \mathbf{H}^u_{\text{cur}} \mathbf{W}^{\text{Q}_1} + \mathbf{b}^{\text{Q}_1}$, $\mathbf{K}_1 = \mathbf{H}^u_{\text{pre}} \mathbf{W}^{\text{K}_1} + \mathbf{b}^{\text{K}_1}$, and $\mathbf{V}_1 = \mathbf{H}^u_{\text{pre}} \mathbf{W}^{\text{V}_1} + \mathbf{b}^{\text{V}_1}$ are the projected behavior representations through linear transformations where $\mathbf{W}^{\text{Q}_1}, \mathbf{W}^{\text{K}_1}, \mathbf{W}^{\text{V}_1} \in \mathbb{R}^{d \times d}$ are weight matrices and $\mathbf{b}^{\text{Q}_1}, \mathbf{b}^{\text{K}_1}, \mathbf{b}^{\text{V}_1} \in \mathbb{R}^{d}$ are bias terms which are introduced to handle the padding vector. To preserve the nature of sequential behaviors, we apply a causal masking by adding the mask matrix $\mathbf{M} \in \mathbb{R}^{N \times N}$ where each element $\mathbf{M}_{ij}$ is defined as follows:
\begin{equation}
\mathbf{M}_{ij} = 
\begin{cases}
0 & \text{if } i \geq j \\
-\infty & \text{if } i < j
\end{cases}
.
\end{equation}
The causal masking ensures that the attention mechanism focuses only on the behaviors at previous time steps and prevents information leakage from future behaviors. Then, the term $\mathrm{softmax} \left(\frac{\mathbf{Q}_1 \mathbf{K}_1^{\mathrm{T}}}{\sqrt{d}}  + \mathbf{M}\right)$ indicates the attention weights that quantify the relevance between the current and each of its previous behaviors. Applying these attention weights to $\mathbf{V}_1$ allows the model to selectively integrate information from previous behaviors that are most relevant to the current behavior. Consequently, the resulting representation $\mathbf{\hat{R}}^u \in \mathbb{R}^{N \times d}$ contains consistent information over user behaviors where $\mathbf{\hat{r}}_t^u\in \mathbb{R}^d$ denotes information about user behaviors from time step $1$ to $t-1$ with considering their relevance with the behavior at time step $t$.

In this step, we apply cross-attention rather than self-attention to reduce the bias that can arise from self-attention as behavior at each time step attends not only to previous steps but also to itself. This bias can overemphasize current behavior and overshadow valuable information from previous behaviors. In contrast, the introduced cross-attention focuses solely on previous time steps, allowing the model to avoid this bias and better capture consistent patterns in user behavior. This approach supports our goal by emphasizing relationships between current and previous behaviors.

\subsubsection{Residual Connection and Layer Normalization}

While the causal cross-attention mechanism effectively reveals consistent patterns by aggregating previous behaviors similar to the current behavior, it excludes the current behavior itself. To address this, we apply a residual connection that reintroduces the current behavior representation $\mathbf{h}_t^u$ into the attention output $\mathbf{\hat{r}}_t^u$ for each time step $t$. By incorporating the residual connection, the model retains crucial information about the current behavior with maintaining the information learned from previous behaviors.

Our goal in this section is to learn a user interest representation by focusing on similar patterns in user behaviors. Since $\mathbf{h}_t^u$ contains both similar and different information compared to $\mathbf{\hat{r}}_t^u$, simply adding them can disrupt the learned patterns. Hence, we add a layer normalization~\cite{ba2016layer} to implicitly emphasize only the similar features of $\mathbf{h}_t^u$. Layer normalization normalizes the representation along the feature dimension, preventing the unique patterns of $\mathbf{h}_t^u$ from being emphasized and enhancing similar patterns between $\mathbf{h}_t^u$ and $\mathbf{\hat{r}}_t^u$. It allows the final representation to capture information of $\mathbf{h}_t^u$ aligned with learned consistent information $\mathbf{\hat{r}}_t^u$. Therefore, the final representation can exhibit the consistent user behavior patterns from time step $1$ to $t$, thereby representing the user's interest.

Overall, the interest representation $\mathbf{R}^u \in \mathbb{R}^{N \times d}$ is derived as follows:
\begin{equation}
\mathbf{R}^u = \mathrm{LayerNorm} \left(\mathbf{\hat{R}}^u + \mathbf{H}^u \right),
\end{equation}
where $\mathbf{r}_t^u \in \mathbb{R}^d$ is the element of $\mathbf{R}^u$ which represents the user's interest representation at time step $t$.

\subsection{Intent Representation Learning}

Based on the view that user behaviors are composed of interests and intentions, we derive the intent representation as the residual obtained by subtracting the interest representation from the behavior representation, denoted as:
\begin{equation}
\mathbf{I}^u = \mathbf{H}^u - \mathbf{R}^u.
\end{equation}
$\mathbf{I}^u = [\mathbf{i}_1^u, \mathbf{i}_2^u, ..., \mathbf{i}_N^u] \in \mathbb{R}^{N \times d}$ represents user intents across all time steps where $\mathbf{i}_t^u\in \mathbb{R}^d$ denotes the user's intent representation at time step $t$. To properly represent user intents, we design an auxiliary loss and model structure that account for both the evolving intent over time and its categorical characteristics.

\subsubsection{Dynamic Intent Representation with Auxiliary Loss}

Intents fluctuate at each time step due to the influence of changing interaction purposes and attributes of each item the user interacts with. However, since multiple interactions can occur with a consistent purpose in a short period, intents can exhibit similar patterns across adjacent time steps. Thus, one key property of intents is continuous changing over time while maintaining similarities between intents at consecutive time steps. In a situation where the model structurally captures the characteristics of interests, intents which are derived by subtracting interests from the behavior representation may indirectly retain some of its characteristics. However, intents are not explicitly modeled within the current architecture.

To directly reflect the dynamic nature of intents, we design a loss function $L_{\text{d}}$ that minimizes the difference between consecutive intent representations by increasing their similarity:
\begin{equation} \label{Ld}
L_{\text{d}} = \frac{1}{N-1} \sum_{t=1}^{N-1} \left(1-\mathrm{sigmoid}\left(\mathrm{sim}\left(\mathbf{i}_t^u, \mathbf{i}_{t+1}^u\right)\right)\right),
\end{equation}
where $\mathrm{sim}(\cdot)$ represents the similarity measure which is determined as the dot product in this study. This loss function allows intents to change gradually across time steps while maintaining coherence between consecutive intents. By introducing this auxiliary loss function, the model is guided to learn not only user interests but also the underlying characteristics of user intents, balancing stability and change over time.

\subsubsection{Categorical Intent Representation with Importance-Weighted Attention}

Users interact with items under various intents which can be viewed as a categorical characteristic of their behaviors. These intent categories represent distinct motivations that influence interactions, meaning that users interact with items based on specific intents belonging to the intent categories. Consequently, intents that are similar across different time steps can be grouped into the same intent category, reflecting a common underlying purpose. In such a situation, rather than focusing on each individual intent representation, the representation of the assigned intent category can reveal a more precise user intent for interactions by indicating a coherent intent category that encompasses all similar intents.
Thus, to learn intent category that will determine the users' next interactions, we assume that users will interact with items under the specific intent category that their most recent intent is assigned. Under this assumption, we enhance the representation of the most recent intent by aggregating similar intents and derive the representation of intent category. This categorization of intents can allow to generalize individual intents and better understand the users' underlying motivations. However, since the detailed relevance of these intents can vary, simple aggregation may not be sufficient. Some intents may exhibit strong unique characteristics specific to their own intent category while others may also be closely related to intents from different categories. To address these differences in relevance, we devise a intent importance extraction module which employs an attention mechanism over the most recent intent and its similar intents to focus on information from the more important intents.

We begin by identifying the intent representations from previous time steps that share similar characteristics with the most recent intent. To achieve this, we compute the similarity between the most recent intent representation $\mathbf{i}_N^u$ and each of the previous intent representations, $\mathbf{i}_1^u$ to $\mathbf{i}_{N-1}^u$. Based on the computed similarities, we filter out the intent representations whose similarity to $\mathbf{i}_N^u$ exceed a predefined threshold. The set of similar intent representations, denoted as $i_{\text{sim}}^u$, is formally defined as:
\begin{equation}
i_{\text{sim}}^u = \left \{ \mathbf{i}_t^u  \mid  \mathrm{sim}(\mathbf{i}_N^u, \mathbf{i}_t^u) \ge \delta, 1 \leq t \leq N-1 \right \} \cup \left \{ \mathbf{i}_N^u \right \},
\end{equation}
where $\delta$ is the threshold that controls the degree of similarity required for intents to be considered similar. The representation matrix for the selected similar intent representations is then defined as $\Tilde{\mathbf{I}}^u = \left[\mathbf{i}_t^u  \mid  \mathbf{i}_t^u \in i_{\text{sim}}^u \right] \in \mathbb{R}^{|i_{\text{sim}}^u| \times d}$.

To learn an intent representation that can accurately reflect its associated intent category, the model should focus on important intents that have strong relevance to other similar intents. Accordingly, motivated by importance extraction module introduced by~\cite{pan2020rethinking}, we compute the relevance scores between intents, denoted as $\mathbf{S} \in \mathbb{R}^{|i_{\text{sim}}^u| \times |i_{\text{sim}}^u|}$, using a modified self-attention mechanism as follows:
\begin{equation}
\mathbf{S} = \mathrm{sigmoid} \left(\frac{\mathbf{Q}_2 \mathbf{K}_2^{\mathrm{T}}}{\sqrt{d}} \right),
\end{equation}
where $\mathbf{Q}_2 = \Tilde{\mathbf{I}}^u\mathbf{W}^{\text{Q}_2}$ and $\mathbf{K}_2 = \Tilde{\mathbf{I}}^u\mathbf{W}^{\text{K}_2}$ are obtained by passing the intent representations through linear transformations. Here, $\mathbf{W}^{\text{Q}_2} \in \mathbb{R}^{d \times d}$ and $\mathbf{W}^{\text{K}_2} \in \mathbb{R}^{d \times d}$ are weight matrices. Instead of using the softmax function which normalizes the relevance score across different intents, we use the sigmoid function to provide an objective measure of the relevance between each pair of intents.

As intent representations that are more relevant to other intents within the same category are important for expressing the intent category, the importance of each intent can be measured by its relevance scores with other intents. To compute the intent importance, we first sum the relevance scores between each intent and other intents whereas excluding itself to avoid high scores between identical intents from distorting the importance computation. We then apply the softmax function to these summed scores to compute the final importance weights for each intent. The importance weight for the $i$-th intent in $i_{\text{sim}}^u$ is computed as follows:
\begin{equation}
\beta_i = \frac{\mathrm{exp} \left(\sum_{j=1, j \neq i}^{|i_{\text{sim}}^u|}\mathbf{S}_{ij}\right)}{\sum_{k=1}^{|i_{\text{sim}}^u|} \mathrm{exp} \left(\sum_{j=1, j \neq k}^{|i_{\text{sim}}^u|}\mathbf{S}_{kj}\right)},
\end{equation}
where $1 \leq i \leq |i_{\text{sim}}^u|$ and $\mathbf{S}_{ij}$ represents the relevance score between the $i$-th and $j$-th intent representations in $i_{\text{sim}}^u$.

The user's final intent representation $\mathbf{i}^u \in \mathbb{R}^d$ is computed as a weighted sum of the selected intent representations where the weights reflect the importance of each intent:
\begin{equation}
\mathbf{i}^u = \sum_{i=1}^{|i_{\text{sim}}^u|} \beta_i \Tilde{\mathbf{i}}_i^u,
\end{equation}
where $\Tilde{\mathbf{i}}_i^u \in \mathbb{R}^d$ is the $i$-th intent representation in $i_{\text{sim}}^u$. This importance-weighted categorical intent representation captures the comprehensive intent of the user at the most recent time step, $N$, by selectively focus on the important intents that strongly indicate the user's categorical intent among the most recent intent and its similar intents from previous time steps. From the aforementioned assumption about user intents at the next time step, $\mathbf{i}^u$ which indicates the intent category of the most recently interacted intent can serve as the intent representation to guide user behaviors for the next interaction.

\subsection{Intent Contrastive Learning}

To further capture the nuanced intent representation, we leverage contrastive learning by focusing on the items which users interact. The interactions which occurred by users' preferences can provide critical insights about their underlying intents. Hence, considering the items that users interact with is essential for understanding their intents. To identify the targets for contrastive learning, we apply sequence augmentation in terms of items by randomly selecting a sequence that shares the same target item as the original sequence. For both the original sequence and the augmented sequence, we derive categorical intent representations. By employing contrastive learning, we aim to refine the intent representation from two perspectives related to the items that users interact with. Basically, the contrastive loss function based on InfoNCE~\cite{oord2018representation} is used to minimize the distance to positive sample to be closer while maximizing the distance to negative samples to be far apart. This loss function for a given anchor $\mathbf{x}$ is defined as follows:
\begin{equation}
L_{\text{CL}} \left(\mathbf{x}, \mathbf{x}_{\text{pos}} \right) = - \mathrm{log} \frac{\mathrm{exp}\left(\mathrm{sim}(\mathbf{x}, \mathbf{x}_{\text{pos}}) / \tau \right)}{\mathrm{exp}\left(\mathrm{sim}(\mathbf{x}, \mathbf{x}_{\text{pos}}) / \tau \right) + \sum_{\mathbf{x}_{\text{n}} \in \mathrm{X}_{\text{neg}}}\mathrm{exp}\left(\mathrm{sim}(\mathbf{x}, \mathbf{x}_\text{n}) / \tau \right)},
\end{equation}
where $(\mathbf{x}, \mathbf{x}_{\text{pos}})$ represents a pair of positive sample representations and $(\mathbf{x}, \mathbf{x}_{\text{n}})$ represents a pair of negative sample representations. $\mathbf{x}_{\text{n}}$ is drawn from the set of negative samples $\mathrm{X}_{\text{neg}}$ and $\tau$ is the temperature parameter.

\subsubsection{Intent-Intent Contrastive Learning}

The augmented sequence $S^{u'}$ is the interaction history of user $u'$ with the target item as $v_{N+1}^u$. The intent representation of this sequence $\mathbf{i}^{u'}$ is derived through the described model architecture. In this context, intent representations for the original and augmented sequences can be treated as positive pairs based on the rationale that users who interact with the same item may share the same intent. Therefore, intent representations for sequences with the same target item should be closer to each other in the latent space.

To achieve this, we apply contrastive learning that brings the intent representations $\mathbf{i}^{u}$ and $\mathbf{i}^{u'}$ closer, while intent representations from sequences with different target items far apart. The contrastive loss function $L_{\text{CL}_1}$ is defined as:
\begin{equation} \label{Lcl1}
L_{\text{CL}_1} = L_{\text{CL}} \left(\mathbf{i}^{u}, \mathbf{i}^{u'} \right) + L_{\text{CL}} \left(\mathbf{i}^{u'}, \mathbf{i}^{u} \right),
\end{equation}
where $(\mathbf{i}^{u}, \mathbf{i}^{u'})$ is a positive pair and $\mathrm{X}_{neg}$ contains intent representations from sequences with different target items. In detail, for a batch size of $|B|$, there are $2|B|$ intent representations after sequence augmentation. Among the $2(|B| - 1)$ potential negative samples in $\mathrm{X}_{\text{neg}}$, some representations might derived from sequences that share the same target item as the positive pair. These are considered as false negative samples and are excluded from $\mathrm{X}_{\text{neg}}$ to ensure that only intent representations from sequences with different target items are included in the negative sample set.

Conversely, users interacting with the same item may have different intentions. Considering the observed relationship between items and intents, this intention can be inferred from the combination of multiple items interacted with under similar intents. While individual items can be associated with various intents, the combination of items can provide a stronger signal about the specific intent behind interactions. Therefore, we directly adjust user intents through contrastive learning with the aggregated representations of items interacted with under intents that constitute the final intent representation.

The intent representation $\mathbf{i}^u$ is derived from the intent representations in $i_{\text{sim}}^u$. Considering these intent representations, we define the item representation matrix $\mathbf{V}_{\text{sim}}^u \in \mathbb{R}^{|i_{\text{sim}}^u| \times d}$ which contains the representations of items interacted with under the intents in $i_{\text{sim}}^u$, as follows:
\begin{equation}
\mathbf{V}_{\text{sim}}^u = [\mathbf{v}_t^u  \mid  t \in \{t \mid \mathbf{i}_t^u \in i_{\text{sim}}^u\}].
\end{equation}
To aggregate these item representations, we simply average them and treat the centroid of the item representations as a representation for the combination of interacted items. The centroid $\bar{\mathbf{v}}^u \in \mathbb{R}^{d}$ is calculated as:
\begin{equation}
\bar{\mathbf{v}}^u = \frac{1}{|\mathbf{V}_{\text{sim}}^u|} \sum_{\mathbf{v}_t^u \in \mathbf{V}_{\text{sim}}^u} \mathbf{v}_t^u.
\end{equation}
This item centroid serves as the positive sample for intent representation for alignment between intent and the items that reflect it. We then apply contrastive learning between the intent representations $\mathbf{i}^u$ and $\mathbf{i}^{u'}$ and the corresponding centroids of their respective item representations $\bar{\mathbf{v}}^u$ and $\bar{\mathbf{v}}^{u'}$. The contrastive loss function $L_{\text{CL}_2}$ is designed to bring each intent representation closer to the centroid of the item representations that were interacted under the corresponding intent as follows:
\begin{equation} \label{Lcl2}
L_{\text{CL}_2} = L_{\text{CL}} \left(\mathbf{i}^{u}, \bar{\mathbf{v}}^{u} \right) + L_{\text{CL}} \left(\bar{\mathbf{v}}^{u}, \mathbf{i}^{u} \right) + L_{\text{CL}} \left(\mathbf{i}^{u'}, \bar{\mathbf{v}}^{u'} \right) + L_{\text{CL}} \left(\bar{\mathbf{v}}^{u'}, \mathbf{i}^{u'} \right).
\end{equation}
Here, contrastive learning is conducted separately for the original sequences and the augmented sequences with $(\mathbf{i}^{u}, \bar{\mathbf{v}}^{u})$ and $(\mathbf{i}^{u'}, \bar{\mathbf{v}}^{u'})$ representing positive pairs for the original and augmented sequences, respectively. We omit false negative removal since the intent representation is learned to be specialized for each user based on their unique interaction history. Hence, all $2(|B| - 1)$ intent and centroid representations from other sequences are treated as negatives and contained in $\mathrm{X}_{\text{neg}}$.

\subsection{Multi-Task Learning}

\subsubsection{User Representation Learning}

For obtained interest representation and intent representation, we aggregate two representations simply by addition to present a user representation. In this way, we assume that the relative contribution of user interest $\mathbf{r}_N^u$ and user intent $\mathbf{i}^u$ are just same on representing user behavior at time step $N$. Then, the final user representation $\mathbf{h}^u \in \mathbb{R}^{d}$ is computed as a combination of the user interest and intent representations as follows:
\begin{equation}
\mathbf{h}^u = \mathbf{r}_N^u + \mathbf{i}^u.
\end{equation}

\subsubsection{Next Item Prediction}

After deriving the user representation, the model predicts how much a user would prefer each item in the candidate item set $\mathcal{V}$. To do this, we compute the dot product between the user representation and the item representation. The user preference score $\hat{y}_i^u$ for item $v_i$ is computed as follows:
\begin{equation}
\hat{y}_i^u = \mathbf{v}_i^{\mathrm{T}} \mathbf{h}^u,
\end{equation}
where $\mathbf{v}_i$ is the item representation for item $v_i$ drawn from $\mathbf{V}$. As before, the dot product works as a similarity measure between the user and the item to indicate how well the item matches the user's preferences. This allows the model to compute preference scores for all items in the candidate set.

For the general recommendation task, the goal of the model is to assign the highest preference score to the ground-truth item $v_g^u$ which represents the item that the user interacted with at the next time step $N+1$. Therefore, we train the model using a cross-entropy loss function which encourages the preference score of the ground-truth item $\hat{y}_g^u$ to be higher than the scores of all other items in $\mathcal{V}$. The training loss $L_{\text{rec}}$ is formulated as follows:
\begin{equation} \label{Lrec}
L_{\text{rec}} = - \frac{\mathrm{exp} (\hat{y}_g^u)}{\sum_{ v_i \in |\mathcal{V}|} \mathrm{exp} (\hat{y}_i^u)}.
\end{equation}

\subsubsection{Model Training}

We train the model using a multi-task learning strategy to jointly optimize three tasks through their corresponding objective functions: the recommendation task via Eq. (\ref{Lrec}), the adjustment task via Eq. (\ref{Ld}), and the contrastive learning tasks via Eq. (\ref{Lcl1}) and Eq. (\ref{Lcl2}). Formally, the total training loss $L$ is defined as a weighted sum of these individual objectives as follows:
\begin{equation}
L = L_{\text{rec}} + \lambda_{\text{d}} L_{\text{d}} + \lambda_{\text{CL}_1} L_{\text{CL}_1} + \lambda_{\text{CL}_2} L_{\text{CL}_2},
\end{equation}
where $\lambda_{\text{d}}$, $\lambda_{\text{CL}_1}$, and $\lambda_{\text{CL}_2}$ are hyperparameters that control the relative contributions of the adjustment and two contrastive learning tasks, respectively.

\section{Experiments} 

\subsection{Experimental Settings}

\subsubsection{Datasets}

We evaluate the effectiveness of our proposed model on three public datasets collected from a real-world platform. Specifically, we utilize the Amazon Review Dataset~\cite{mcauley2015image, he2016ups} which presents user review data collected from Amazon which is one of the largest e-commerce platform. This dataset can be divided into various subsets based on item categories. Following many existing studies in contrastive learning-based recommendation and intent-based recommendation~\cite{zhou2020s3, qin2023meta, chen2022intent, qin2024intent}, in this paper, we adopt three representative subcategories: “Sports and Outdoors”, “Beauty”, and “Toys and Games”.

For high quality of datasets, we preprocess the datasets following common practice in~\cite{xie2022contrastive, qiu2022contrastive, chen2022intent}. All interactions are considered as implicit feedback which treat all observed interactions as positive feedback while others as negative regardless of explicit numeric ratings. Then, we keep the “5-core” datasets which eliminate users and items with less than five interactions. For model training and evaluation, we divide the datasets into training, validation, and test datasets. Specifically, we chronologically sort the historical items for each user according to the interaction timestamps to construct the user's interaction sequence. Following recent studies that contain data augmentation through sequence splitting in implementation and validate its effectiveness~\cite{tan2016improved, zhou2024contrastive, qiu2022contrastive, li2023multi, qin2024intent}, we split each user sequence into subsequences and augment the data with these additional subsequences. For each user sequence, the last interacted item is used as the target for test where the remaining interactions are considered as the input sequence. Similarly, the second-to-last item is used for validation and the other items are used for training. Table~\ref{dataset} summarizes the statistical details of these datasets after preprocessing.

%\captionsetup[table]{skip=5pt}
\begin{table}[b]
\centering
\caption{Statistics of the datasets.} \label{dataset}
\resizebox{0.99\linewidth}{!}{%
\begin{tabular}{c|ccccc}
\hline
Dataset & Num.Users & Num.Items & Num.Interactions & Avg.Length & Sparsity \\ \hline\hline
Sports & 35,598 & 18,357 & 296,337 & 8.3 & 99.95\% \\
Beauty & 22,363 & 12,101 & 198,502 & 8.9 & 99.93\%   \\
Toys & 19,412 & 11,924 & 167,597 & 8.6 & 99.93\%  \\  \hline
\end{tabular}%
}
\end{table}

\subsubsection{Evaluation Metrics}

To ensure a fair model evaluation, we predict the user preference scores for items in the whole item set without negative sampling~\cite{krichene2020sampled, wang2019neural}. Then, the performance for the model prediction is evaluated through top-$k$ Hit Ratio (HR@$k$) and Normalized Discounted Cumulative Gain (NDCG@$k$)~\cite{jarvelin2002cumulated} which are commonly used evaluation metrics in top-$k$ recommendation. In short, HR measures whether the ground-truth item presents in top-$k$ item list derived by model and NDCG measures how well the predicted order of items aligns with the actual order of items in user sequence. We report HR and NDCG with $k \in {5, 10, 20}$.

\subsubsection{Baselines}

To verify the effectiveness of our model, we compare our model with the following representative baselines belonging to three groups of sequential recommendations:

\begin{itemize}
\item \textbf{General sequential models}: GRU4Rec~\cite{hidasi2015session} applies gated recurrent unit to model user sequences. SASRec~\cite{kang2018self} employs an uni-directional self-attention mechanism to capture sequential patterns in user sequences, becoming one of the SOTA model in sequential recommendation task.
\item \textbf{Sequential models with self-supervised learning}: BERT4Rec~\cite{sun2019bert4rec} adopts a bi-directional self-attention mechanism and performs a Cloze task which predicts masked items instead of the next item, inspired by BERT~\cite{devlin2018bert}. S\textsuperscript{3}-Rec\textsubscript{MIP}~\cite{zhou2020s3} introduces self-supervised learning to capture correlations between items, attributes, and sequences. Since all our model and other baselines do not consider item attribute information, we only contain masked item prediction module for fair comparison, namely S\textsuperscript{3}-Rec\textsubscript{MIP}. CL4SRec~\cite{xie2022contrastive} proposes three data augmentation methods and first applies contrastive learning between original and augmented sequences for sequential recommendation task. CoSeRec~\cite{liu2021contrastive} further devises two data augmentation strategies based on item correlations for robustness of augmented sequence. DuoRec~\cite{qiu2022contrastive} introduces an unsupervised model-level augmentation method using dropout and a supervised positive sampling strategy that accounts for semantic similarity between sequences.
\item \textbf{Sequential models considering intent}: DSSRec~\cite{ma2020disentangled} proposes a sequence-to-sequence training strategy that contains self-supervision and intent disentanglement. ICLRec~\cite{chen2022intent} performs clustering on intent representations and contrastive learning that reflects the learned intent clusters for latent intent modeling. IOCRec~\cite{li2023multi} designs intent-level contrastive learning to emphasize the main intent among multiple intents for user behavior. ICSRec~\cite{qin2024intent} further improves ICLRec by leveraging data augmentation considering subsequences and sequence augmentation considering target items.
\end{itemize}

\subsubsection{Implementation Details}

We implement BERT4Rec, S\textsuperscript{3}-Rec\textsubscript{MIP}, CoSeRec, DuoRec, ICLRec, IOCRec, and ICSRec using official codes provided by authors. GRU4Rec, SASRec, CL4SRec, and DSSRec are implemented based on public resources. For all baselines, we follow the optimal hyperparameter combinations reported in their original papers. Our model is implemented in PyTorch. For IDCLRec, the sequence encoder consists of 2 number of self-attention blocks and attention heads. The embedding dimension $d$ is set as 64 and the maximum sequence length $N$ is 50. We apply a dropout rate of 0.5 and set the temperature parameter $\tau$ as 1. The model is optimized using Adam optimizer~\cite{kingma2014adam} with a learning rate of 0.001 and a batch size of 256. The model is trained for up to 300 epochs with applying early stopping which stops training if the NDCG@20 on the validation dataset does not improve for 40 consecutive epochs. For hyperparameter tuning, we search for the best combination of the intent similarity threshold $\delta$ within \{0.5, 0.6, 0.7, 0.8, 0.9\} and weights for each loss component $\lambda_{\text{d}}$, $\lambda_{\text{CL}_1}$, and $\lambda_{\text{CL}_2}$ within \{0.1, 0.2, 0.3, 0.4, 0.5\} through grid search. All experiments are conducted using three different random seeds and the average results are reported to ensure robustness of models.

\subsection{Performance Comparison}

\begin{table*}[t]
\renewcommand{\arraystretch}{1.1}
\centering
\caption{Performance comparisons of the proposed model with baselines where best results are boldfaced and second-best results are underlined.}
\label{comparison}
\resizebox{\textwidth}{!}{%
\begin{tabular}{cc|cc|ccccc|cccc|c}
\hline
Dataset & Metric & GRU4Rec & SASRec & BERT4Rec & S\textsuperscript{3}-Rec\textsubscript{MIP} & CL4SRec & CoSeRec & DuoRec & DSSRec & ICLRec & IOCRec & ICSRec & IDCLRec \\ \hline\hline
Sports & HR@5 & 0.0101 &  0.0191 & 0.0226 & 0.0119 & 0.0246 & 0.0271 & 0.0326 & 0.0195 & 0.0274 & 0.0294 & \underline{0.0383} & \textbf{0.0431} \\
 & HR@10 & 0.0176 & 0.0300 & 0.0354 & 0.0199 & 0.0389 & 0.0404 & 0.0502 & 0.0309 & 0.0425 & 0.0467 & \underline{0.0549} & \textbf{0.0618} \\
 & HR@20 & 0.0293 &  0.0462 & 0.0536 & 0.0326 & 0.0596 & 0.0599 & 0.0744 & 0.0461 & 0.0640 & 0.0692 & \underline{0.0766} & \textbf{0.0878} \\
 & NDCG@5 & 0.0063 & 0.0125 & 0.0151 & 0.0076 & 0.0162 & 0.0183 & 0.0192 & 0.0130 & 0.0179 & 0.0168 & \underline{0.0270} & \textbf{0.0299} \\
 & NDCG@10 & 0.0087 & 0.0160 & 0.0192 & 0.0102 & 0.0207 & 0.0226 & 0.0249 & 0.0167 & 0.0227 & 0.0223 & \underline{0.0323} & \textbf{0.0360} \\
 & NDCG@20 & 0.0117 & 0.0201 & 0.0237 & 0.0134 & 0.0260 & 0.0275 & 0.0310 & 0.0205 & 0.0281 & 0.0279 & \underline{0.0378} & \textbf{0.0426} \\ \hline
Beauty & HR@5 & 0.0169 & 0.0349 & 0.0390 & 0.0172 & 0.0396 & 0.0482 & 0.0590 & 0.0391 & 0.0504 & 0.0515 & \underline{0.0677} & \textbf{0.0714} \\
 & HR@10 & 0.0296 & 0.0545 & 0.0602 & 0.0300 & 0.0636 & 0.0712 & 0.0875 & 0.0615 & 0.0727 & 0.0816 & \underline{0.0932} & \textbf{0.1013} \\
 & HR@20 & 0.0487 & 0.0843 & 0.0900 & 0.0454 & 0.0950 & 0.1032 & 0.1243 & 0.0918 & 0.1055 & 0.1179 & \underline{0.1284} & \textbf{0.1399} \\
 & NDCG@5 & 0.0105 & 0.0223 & 0.0251 & 0.0106 & 0.0255 & 0.0319 & 0.0363 & 0.0255 & 0.0331 & 0.0306 & \underline{0.0484} & \textbf{0.0507} \\
 & NDCG@10 & 0.0145 & 0.0286 & 0.0319 & 0.0146 & 0.0332 & 0.0393 & 0.0455 & 0.0327 & 0.0403 & 0.0404 & \underline{0.0566} & \textbf{0.0604} \\
 & NDCG@20 & 0.0193 & 0.0361 & 0.0394 & 0.0186 & 0.0410 & 0.0473 & 0.0548 & 0.0403 & 0.0485 & 0.0495 & \underline{0.0655} & \textbf{0.0701} \\ \hline
Toys & HR@5 & 0.0173 & 0.0438 & 0.0320 & 0.0090 & 0.0416 & 0.0565 & 0.0668 & 0.0493 & 0.0573 & 0.0579 & \underline{0.0766} & \textbf{0.0832} \\
 & HR@10 & 0.0297 & 0.0642 & 0.0490 & 0.0168 & 0.0631 & 0.0800 & 0.0958 & 0.0709 & 0.0818 & 0.0872 & \underline{0.1032} & \textbf{0.1129} \\
 & HR@20 & 0.0491 & 0.0918 & 0.0733 & 0.0296 & 0.0913 & 0.1089 & 0.1307 & 0.0972 & 0.1134 & 0.1210 & \underline{0.1348} & \textbf{0.1498} \\
 & NDCG@5 & 0.0109 & 0.0291 & 0.0211 & 0.0054 & 0.0279 & 0.0391 & 0.0396 & 0.0336 & 0.0393 & 0.0331 & \underline{0.0560} & \textbf{0.0595} \\
 & NDCG@10 & 0.0149 & 0.0357 & 0.0266 & 0.0079 & 0.0348 & 0.0467 & 0.0490 & 0.0406 & 0.0472 & 0.0426 & \underline{0.0646} & \textbf{0.0690} \\
 & NDCG@20 & 0.0197 & 0.0426 & 0.0327 & 0.0111 & 0.0419 & 0.0540 & 0.0578 & 0.0472 & 0.0551 & 0.0511 & \underline{0.0726} & \textbf{0.0784} \\ \hline
\end{tabular}%
}
\end{table*}

Table \ref{comparison} shows the performance of the proposed model and all baselines across three datasets. In summary, our proposed IDCLRec significantly outperforms other baselines on all metrics and datasets.

Specifically, for general sequential models, SASRec shows better performance than GRU4Rec suggesting that the attention mechanism in SASRec can capture sequential patterns in user behaviors more effectively than RNN-based GRU4Rec. Although BERT4Rec adopts a self-supervised learning strategy in addition to the attention mechanism, it sometimes performs better or worse than SASRec. These inconsistent results might arise as the masked item prediction in BERT4Rec does not fit well with the sequential recommendation task that aims to predict the next item whereas the bi-directional attention mechanism well captures behavior dependencies. S\textsuperscript{3}-Rec\textsubscript{MIP} which also employs masked item prediction performs poorly in all situations. While S\textsuperscript{3}-Rec includes several modules that utilize additional item attribute information, we do not consider these modules to exclude attribute information. In addition to the conflict on item prediction task, this model modification may negatively impact the model's effectiveness. Moreover, CL4SRec, CoSeRec, and DuoRec achieve performance improvements over other self-supervised learning-based models by incorporating contrastive learning to enhance sequence representations. CoSeRec improves CL4SRec by augmenting sequences based on item correlations and DuoRec further outperforms CoSeRec by considering semantic similarities between sequences during augmentation. These results for general sequential models and models with self-supervised learning highlight the effectiveness of the transformer encoder architecture with attention mechanisms and the advantages of introducing contrastive learning for representation learning.

For intent-based sequential models, they achieve superior performance compared to other baselines by interpreting the representations of the user interaction sequences as user intent from the user behavioral perspective. While the performance of DSSRec exceeds the performance of SASRec, ICLRec, IOCRec, and ICSRec surpass DSSRec by enhancing intent representations through contrastive learning. ICLRec and IOCRec demonstrate comparable performance with employing different contrastive learning strategy. ICLRec captures contrast signals between different intents and between intents and their associated clusters, whereas IOCRec focuses on capturing contrast signals between multiple intents. DuoRec and ICSRec consistently outperform these models and show the best performance among the baselines. Unlike other models that utilize contrastive learning with sequence augmentation which modifies the original sequence by altering item composition, they determine the sequence having the same target item with the original sequence as the augmented sequence. Additionally, they augment the training dataset itself through a splitting operation for model training. Overall, learning sequence representations by giving them the meaning of user intent from the perspective of user behavior is shown to be more effective in understanding users' preferences compared to learning without attributing specific semantics to sequences. Also, results indicate that sequence augmentation based on the target item can effectively generate sequences with similar context and data augmentation through sequence splitting can enrich the training process by expanding the dataset.

Finally, IDCLRec demonstrates the best performance across all baselines under all evaluation scenarios. Specifically, IDCLRec achieves 5.47\% to 14.62\% and 4.75\% to 12.70\% performance improvements over the best baseline in HR and NDCG, respectively. These results confirm the effectiveness of our proposed model. In detail, both HR and NDCG consistently exhibit highly improved performance. The improvement in HR suggests that IDCLRec can effectively identify relevant items within the recommendation lists by accurately capturing user preferences. Similarly, the improvement in NDCG implies that IDCLRec can retrieve the correct ranking of items by prioritizing more relevant items higher in the list. This consistent improvement indicates the model capability of understanding preferable items and their relative importance. Furthermore, the average performance improvement in the Sports dataset is 12.44\% which is significantly higher than the 6.93\% in Beauty and 8.37\% in Toys. Where Sports is little sparser than other datasets, this degree of overall improvement suggests that our model can excel in handling data sparsity with capturing subtle user intent and user preferences. Regarding the development of discovered beneficial components by addressing overlooked points, we can conclude key factors of IDCLRec which contribute to overall performance improvements as follows: 1) Beyond merely considering user intents within user behaviors, our approach which explicitly explore both user interests and intents enables more comprehensive user behavior modeling. 2) The introduction of a contrastive loss function appropriate for the intent perspective with accounting for the characteristics of user intent facilitates more precise intent representation learning.

\subsection{Ablation Study}

\begin{table}[t]
\centering
\caption{Ablation study of the proposed model.}
\label{ablation}
\resizebox{\linewidth}{!}{%
\begin{tabular}{c|cc|cc|cc}
\hline
{Model} & \multicolumn{2}{c|}{Sports} & \multicolumn{2}{c|}{Beauty} & \multicolumn{2}{c}{Toys} \\ \cline{2-3} \cline{4-5} \cline{6-7}
 & HR@20 & NDCG@20 & HR@20 & NDCG@20 & HR@20 & NDCG@20 \\ \hline\hline
IDCLRec & \textbf{0.0878} & \textbf{0.0426} & \textbf{0.1399} & \textbf{0.0701} & \textbf{0.1498} & \textbf{0.0784} \\ \hline
(A) w/o intent difference minimization & 0.0867 & 0.0425 & 0.1386 & 0.0695 & 0.1476 & 0.0776   \\
(B) w/o intent-intent contrastive learning & 0.0837 & 0.0407 & 0.1377 & 0.0693 & 0.1457 & 0.0766   \\
(C) w/o intent-item contrastive learning & 0.0852 & 0.0410 & 0.1361 & 0.0684 & 0.1427 & 0.0758   \\ %\hline
(D) w/o all auxiliary losses & 0.0806 & 0.0379 & 0.1347 & 0.0664 & 0.1386 & 0.0712   \\ \hline
(E) w/o intent-interest disentanglement & 0.0778 & 0.0355 & 0.1325 & 0.0633 & 0.1335 & 0.0656   \\
(F) w/o intent categorization & 0.0799 & 0.0394 & 0.1309 & 0.0669 & 0.1401 & 0.0749   \\
(G) w/o importance-weighted attention & 0.0859 & 0.0418 & 0.1368 & 0.0691 & 0.1474 & 0.0776   \\ \hline
\end{tabular}%
}
\end{table}

To investigate the effectiveness of each component of our proposed model, we conduct an ablation study on seven variants of IDCLRec in terms of training loss and model architecture. The four variants focusing on the effect of different training losses are detailed as follows: (A) removes the minimization of consecutive intent differences by setting $L_{\text{d}}$ as 0, (B) removes intent-intent contrastive learning by setting $L_{\text{CL}_1}$ as 0, (C) removes intent-item contrastive learning by setting $L_{\text{CL}_2}$ as 0, and (D) removes all auxiliary losses by setting $L_{\text{d}}$, $L_{\text{CL}_1}$, and $L_{\text{CL}_2}$ as 0. The other three variants focusing on the effect of model architecture, applying all auxiliary losses to the intents derived in each case, are as follows: (E) treats whole behaviors as only intents without disentanglement of intents and interests in behaviors by setting $\mathbf{H}^u = \mathbf{I}^u$, (F) uses the most recent intent instead of the categorized intent by setting $\mathbf{i}^u = \mathbf{i}_N^u$, and (G) uses average pooling instead of the importance-weighted attention mechanism for intent importance extraction by setting $\mathbf{i}^u = \sum_{i=1}^{|i_{\text{sim}}^u|} \frac{1}{|i_{\text{sim}}^u|} \Tilde{\mathbf{i}}_i^u$. Table~\ref{ablation} reports HR@20 and NDCG@20 for the proposed model and its variants. Overall, the results clearly reveal that IDCLRec consistently outperforms all other variants, indicating the effectiveness of the proposed components in enhancing the model's ability to capture user preferences accurately.

\subsubsection{Effect of Training Loss}

Comparing IDCLRec with variants (A), (B), and (C), we observe that the removal of any loss function results in a decrease in performance. This implies that all the proposed loss functions are essential for improving the recommendation quality, while the importance of each depends on the degree of performance degradation. For (A), the performance decrease suggests that constraining intents to be closer in consecutive time steps helps to capture the user intents behind the users' sequential behaviors. However, the relatively small performance drop could be explained by the fact that the characteristics of intents might be implicitly contained without explicitly aligning consecutive intents. This might be because intents are expressed by subtracting the learned interest representations reflecting its properties from the behavior representations which consists of interests and intents, thereby residual intents preserve some of the underlying intent properties. The performances for (B) and (C) show that both types of contrastive learning loss functions contribute positively to express complicated intents. Specifically, (C) shows more performance degradation in Beauty and Toys while (B) is more poor in Sports. From this results, we can check that the intent enhancement through contrastive learning between intents and items is more significant in Beauty and Toys since item combinations might allow to observe specific intents although each item interaction could be occurred under multiple intents. In contrast, Sports shows more enhancement through contrastive learning between intents, implying that each item in Sports might be strongly associated with a particular intent. By the way, in different situations, we can conclude that modeling intents in relation to item combinations or single item itself is beneficial as the model can directly connect user intents to interacted items. Furthermore, the superior performance of (A) compared to (B) and (C) suggests that integration of both proposed contrastive learning losses can have a synergistic effect in helping to express user intent more comprehensively.

In variant (D) that excludes all loss functions in model training, the performance is worse than variants which remove only one of each loss function. It can be seen that all the proposed loss functions work together in a positive direction for performance improvement without distorting the user representation. When comparing IDCLRec, (D), and ICSRec which is SOTA baseline, we observe that (D) achieves consistently higher HR than SOTA baseline, although its NDCG is lower. However, in IDCLRec that is enhanced form of (D) by training with all three proposed losses, NDCG substantially improves and ultimately reaches the best performance. A distinctive feature of (D) is its model architectural approach: 1) disentangles user behaviors into intents and interests and 2) enhances the most recent intent into the intent category it belongs to by grouping intents at each user-level. The higher HR compared to the SOTA baseline indicates that this approach is particularly effective in finding items that match user preferences. Furthermore, the three loss functions are specifically designed to capture different characteristics of intents and facilitate a nuanced adjustment of intents. The significant increase in NDCG when incorporating all proposed losses in IDCLRec validates that careful adjustment of intent representations through these losses leads to more precise ranking of preferred items, beyond merely finding them. These detailed observations on each evaluation metric demonstrate the architectural and training advantages of the proposed model in both identifying and accurately ranking items according to user preferences.

\subsubsection{Effect of Model Architecture}

For variant (E), the result reveals remarkably lower performance compared to IDCLRec and even (D) despite being trained with all auxiliary loss functions. This sharp performance drop highlights the importance of viewing user behaviors as combination of intents and interests to thoroughly interpret user behaviors and provide accurate recommendations. In detail, we can verify that viewing entire behaviors as intents as in existing studies leads to unclear exploration of the behaviors due to oversimplified generalizations. Where the losses are designed and applied based on the characteristics of intents, (E) fails to distinguish between intents and interests. Thus, by simplifying behaviors which inherently include both intents and interests exclusively as intents and applying intent-specific losses to them, the representations fail to disentangle distinct information, leading to entangled and ambiguous representations that degrade the overall quality of the recommendation. This observation underscores that user behaviors are not solely composed of intents but consist of both intents and interests that possess distinct characteristics that require tailored representation learning based on their unique properties.

The performance degradation from (F) suggests that reflecting only the intent for a single interaction is insufficient to robustly represent users' current purposes. Instead, aggregating similar intents across multiple interactions to learn user-wise categorical intents is more powerful for capturing user intents. Particularly, the noticeable drop in HR compared to NDCG implies that reflecting only the most recent intent provides overly strong signals confined to a specific interaction. This lack of generalizability for intentions may limit the ability to seek for diverse candidate items that align with subsequent interactions. The proposed intent categorization in IDCLRec achieves improvements by considering how categorical characteristics of intents emerge from user behaviors and adapting these characteristics from the perspective of each individual user. By doing so, it can effectively replace previous approaches that arbitrarily divide all users' intents into a fixed number of intent categories. Moreover, the impact of integrating intents about multiple interactions aligns with the observed effectiveness of learning intents through item combinations associated with similar intents as mentioned in (C). Finally, the result for (G) confirms that accounting for the varying importance of different intents within the same intent category yields a performance improvement. The higher performance of (G) compared to (F) indicates that all similar intents can collectively represent an intent category to some extent and contribute to offer a list of items that match the intention for the next interaction. However, the superior performance of IDCLRec implies that certain intents can play a more crucial role in defining their assigned intent category. Therefore, weighting each intent according to its relevance within the assigned category enables to generate more accurate intent representations.

\subsection{Hyperparameter Sensitivity Analyses}

\begin{figure}[t]
\centering
\includegraphics[width=0.99\linewidth]{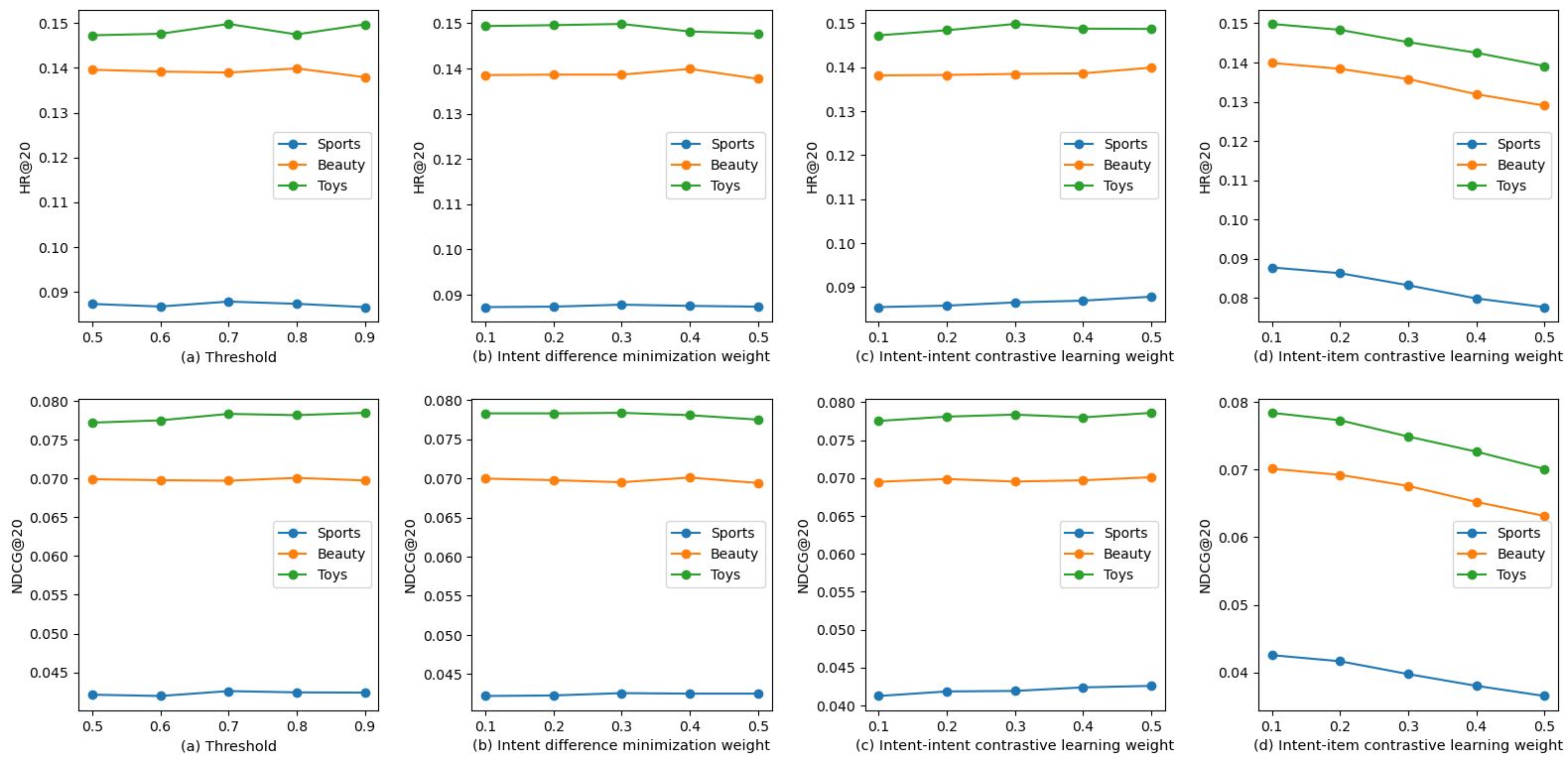} 
\caption{Performance comparisons of the proposed model on different hyperparameters.} \label{hyperparameter}
\end{figure}

Our model training involves four hyperparameters. To assess the sensitivity of each hyperparameter, we implement experiments by setting five candidate values for each. For each experiment, we change only value of target hyperparameter at a time, keeping the others fixed at their optimal values. The results for HR@20 and NDCG@20 across different hyperparameter combinations are shown in Figure \ref{hyperparameter}.

\subsubsection{Impact of Threshold}

The threshold $\delta$ defines the similarity criterion for selecting relevant intent representations when representing categorical intent. Figure \ref{hyperparameter}-(a) is results for five threshold candidates \{0.5, 0.6, 0.7, 0.8, 0.9\}. The optimal threshold value is found to be 0.7 for Sports and Toys and 0.8 for Beauty. Where there are little performance differences from different value of thresholds, we can think that the results are similar across all candidates. When the threshold is lower, more intents including those with lower similarity are considered to capture the intent category, resulting in a more inclusive representation. However, through the proposed importance-weighted attention mechanism, intent representations with lower similarity to others within the same intent category are naturally assigned less weight. This allows highly similar intents to dominate the final categorical intent representation. Consequently, the final representation can be close to representation formed by only high-similarity intents selected from higher thresholds. Moreover, this understanding about the role of importance-weighted attention can implicitly explain the observed performance drops when average pooling is used instead of attention in the ablation study. To further verify this observation regarding the consistency of the results, we examine the case when the threshold is set to 0 which means all intents are considered for categorical intent learning. In this scenario, the performance is almost identical to the performance when the threshold is 0.5 with HR differing by less than 0.0007 and NDCG varying by less than 0.0001 across all datasets. These results can be explained by two factors. First, since similarity is calculated as the sigmoid of a dot product, very few intent pairs may actually have similarity below 0.5. Thus, including all intents or only those above a similarity threshold of 0.5 may yield almost the same set of intents. Second, these results also highlight the effectiveness of the importance-weighted attention that can naturally mitigate the influence of less similar intents regardless of whether a threshold is applied. Overall, while observed optimal thresholds can output the best performances, given that performances remain stable across various thresholds, introducing an additional hyperparameter that is threshold may be unnecessary. Instead, we can consider all intents for the intent importance extraction module by relying on the proposed attention mechanism to effectively emphasize the most relevant intents.

\subsubsection{Impact of Intent Difference Minimization Weight}

The intent difference minimization loss weight $\lambda_{\text{d}}$ controls the degree of the loss function $L_{\text{d}}$ that captures the dynamic nature of intents throughout training. We verify the impact of $\lambda_{\text{d}}$ with five candidates \{0.1, 0.2, 0.3, 0.4, 0.5\} as displayed in Figure \ref{hyperparameter}-(b). The best performance is achieved when weight is 0.3 for Sports and Toys and 0.4 for Beauty. In general, the variations in weight do not result in remarkable performance differences, demonstrating that the model maintains robust performance across different weight values. The mere reflection of $L_{\text{d}}$ can effectively encourage similar intents for adjacent user interactions. This low sensitivity might also explain an implicit reflection of the dynamic characteristics of intents within the originally extracted intent representations.

\subsubsection{Impact of Intent-Intent Contrastive Learning Weight}

The intent-intent contrastive learning loss weight $\lambda_{\text{CL}_1}$ controls the degree of the loss function $L_{\text{CL}_1}$ that aligns intents for interacting same items. For five candidates \{0.1, 0.2, 0.3, 0.4, 0.5\}, 0.5 is optimal value for Sports and Beauty and 0.3 is optimal for Toys. As shown in Figure \ref{hyperparameter}-(c), there is some sensitivity to $\lambda_{\text{CL}_1}$ where overall performance improves as $\lambda_{\text{CL}_1}$ increases. Specifically, for Sports and Beauty, performance consistently improves with higher values. For Toys, while 0.3 is selected to be optimal value, 0.4 and 0.5 yield comparable results. Hence, we can conclude that higher values of $\lambda_{\text{CL}_1}$ generally lead to better performance across datasets.

\subsubsection{Impact of Intent-Item Contrastive Learning Weight}

The intent-item contrastive learning loss weight $\lambda_{\text{CL}_2}$ controls the degree of the loss function $L_{\text{CL}_2}$ that aligns intents and corresponding item combinations. We conduct experiments with five candidates \{0.1, 0.2, 0.3, 0.4, 0.5\} and the results are illustrated in Figure \ref{hyperparameter}-(d). Across all datasets, the best performances are consistently observed when the value is set to 0.1 that is the smallest value among the candidates. $\lambda_{\text{CL}_2}$ exhibits high sensitivity with performance deteriorating significantly as the value increases unlike other hyperparameters.

Regarding the observed impact of each loss weight, we further interpret the results in terms of the relationship between loss weights and model performances. Related to the ablation study results, there are two interesting insights about loss weights and performances for contrastive learning tasks: 1) Although the optimal weight for intent-item contrastive learning is small as 0.1, intent-item contrastive learning loss plays a critical role in model performance by showing a significant performance drop when removing it. 2) Despite intent-intent contrastive learning requires higher optimal weights compared to intent-item, the ablation results indicate that adding intent-item contrastive learning generally leads to a larger performance improvement. Loss weights are used to balance the influence of different loss functions during model training. Although the intent-item contrastive learning loss has smaller assigned weight compared to other losses, its presence is crucial in guiding the model toward better performance. Also, a higher loss weight does not necessarily imply a greater impact on model performance. In this point, we can think that the effectiveness of each loss function depends not only on their assigned weights but also on the quality and uniqueness of the information it provides. The intent-item contrastive learning loss may offer valuable insights that other tasks cannot cover so that removing this loss can lead to a more performance degradation due to the absence of such unique characteristics. In contrast, the relatively less performance drop from the removal of the intent-intent contrastive learning loss may suggests that it provides some redundant information. From this perspective, we can think that the effectiveness of a loss function is determined more by the information it provides than by its assigned weight in this study. Thus, despite its smaller optimal weight, intent-item contrastive learning can yield a more performance improvement than intent-intent contrastive learning. We can infer that adjustment of intents through multiple item combinations can capture essential patterns of intent representations for better recommendations whereas adjustment through a single item might learn less critical information.

\subsection{Intent-Interest Analysis}

\begin{figure}[t]
\centering
\includegraphics[width=0.85\linewidth]{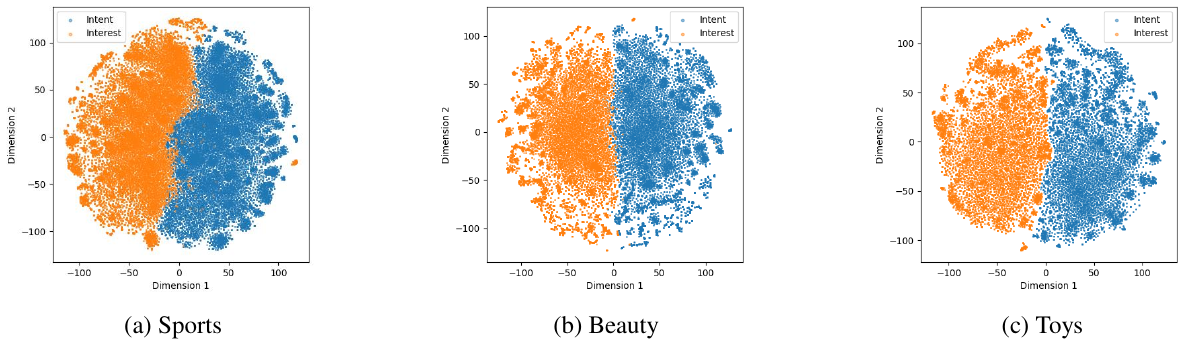} 
\caption{Visualization of intent and interest representations.} \label{tsne}
\end{figure}

To investigate the distinction between intents and interests extracted from user behaviors, we visualize these two features using t-SNE~\cite{van2008visualizing}. As shown in Figure \ref{tsne}, the intent and interest representations of each user at their most recent time step are projected onto a 2D space. The resulting visualization reveals a strikingly clear separation between intent and interest representations with minimal overlap. This separation demonstrates that intents and interests are inherently different user characteristics. Intents, reflecting motivational short-term behaviors, are distinctly localized compared to interests which encapsulate more personal long-term behaviors. The lack of significant overlap between the two distributions highlights their complementary roles in characterizing user behaviors. Moreover, the result emphasizes the effectiveness of our proposed disentanglement approach, particularly the causal cross-attention mechanism, in capturing and isolating these distinct behavioral factors. Overall, the visualization provides compelling evidence of the distributional independence of intents and interests and supports the efficacy of our method in modeling each intent and interest considering their unique properties.

\section{Conclusion} 

In this paper, we propose IDCLRec, a novel model that captures user behaviors more accurately by analyzing the nuanced semantics and characteristics underlying the behaviors that yield user-item interactions. Unlike previous studies that interpret user behaviors solely based on intents or without semantic frameworks, IDCLRec disentangles behaviors into intents and interests. We extract interests through a causal cross-attention mechanism that reflects the consistent characteristics of interest over time, whereas deriving intents through an additional loss function and importance-weighted attention that account for the dynamic and categorical characteristics of intent. Moreover, we introduce two types of contrastive learning loss functions that address different relationships between intents and items to better understand complex user intents. Through this modeling, IDCLRec achieves a more refined understanding of user behaviors. Experiments on real-world datasets demonstrate the effectiveness of IDCLRec through significant performance improvements.

\bibliographystyle{ACM-Reference-Format}
\bibliography{reference}

\end{document}